%% file: main.tex
\documentclass[sigplan,10pt,nonacm]{acmart}
\renewcommand\footnotetextcopyrightpermission[1]{}
\usepackage{arydshln}
\usepackage{enumitem}
\usepackage{subcaption}
\usepackage{xspace}
\AtBeginDocument{%
  }

\newenvironment{packeditemize}{
\begin{itemize}
  \setlength{\itemsep}{0.3pt}
  \setlength{\parskip}{2pt}
  \setlength{\parsep}{0pt}
}{\end{itemize}}

\newcommand{\sys}{RecoGEM\xspace}

\begin{document}

\title{RecoGEM: A Serving System for Production-Scale Generative Recommendation}
\author{Xiangrui Yang}
\authornote{Both authors contributed equally to this research.}
\affiliation{%
  \institution{The University of Hong Kong}
  \city{Hong Kong}
  \country{China}
}
\email{yxr620@gmail.com}

\author{Yunfeng Zhao}
\authornotemark[1]
\affiliation{%
  \institution{Kuaishou Technology}
  \city{Beijing}
  \country{China}
}
\email{zhaoyunfeng@kuaishou.com}

\author{Jiawei Yang}
\authornote{Work done while at Kuaishou Technology.}
\affiliation{%
  \institution{Peking University}
  \city{Beijing}
  \country{China}
}
\email{yangjiawei26@stu.pku.edu.cn}

\author{Fangyu Zhang}
\affiliation{%
  \institution{Kuaishou Technology}
  \city{Beijing}
  \country{China}
}
\email{zhangfangyu@kuaishou.com}

\author{Enrui Zhao}
\affiliation{%
  \institution{Renmin University of China}
  \city{Beijing}
  \country{China}
}
\email{zhaoenrui@ruc.edu.cn}

\author{Liang Zeng}
\affiliation{%
  \institution{Kuaishou Technology}
  \city{Beijing}
  \country{China}
}
\email{zengliang03@kuaishou.com}

\author{Cheng Peng}
\affiliation{%
  \institution{Kuaishou Technology}
  \city{Beijing}
  \country{China}
}
\email{pengcheng09@kuaishou.com}

\author{Minhui Xie}
\affiliation{%
  \institution{Renmin University of China}
  \city{Beijing}
  \country{China}
}
\email{xieminhui@ruc.edu.cn}

\author{Jiaqiang Liu}
\authornote{Corresponding author.}
\affiliation{%
  \institution{Kuaishou Technology}
  \city{Beijing}
  \country{China}
}
\email{liujiaqiang@kuaishou.com}

\author{Xiao Liang}
\affiliation{%
  \institution{Kuaishou Technology}
  \city{Beijing}
  \country{China}
}
\email{liangxiao03@kuaishou.com}

\author{Yiming Qiu}
\affiliation{%
  \institution{The University of Hong Kong}
  \city{Hong Kong}
  \country{China}
}
\email{yimingq@hku.hk}

\renewcommand{\shortauthors}{Xiangrui Yang et al.}

\renewcommand\footnotetextcopyrightpermission[1]{}

\begin{abstract}

Generative recommendation (GR) is moving industrial recommender systems (RecSys) from scoring candidate items to autoregressively generating item identifiers.
This shift promises higher recommendation quality, but makes serving substantially harder: each request executes a short sequence of dependent decoding steps that explore a large item space and select promising candidates on the latency-critical path.
GR not only creates structural mismatches with conventional RecSys pipelines, but also challenges the architectural and optimization assumptions of LLM serving stacks, making direct adoption of existing systems difficult.

We present \sys, our production GR serving system.
\sys uses a runtime--engine decoupling architecture to move GR-specific serving responsibilities, such as candidate-state transitions and live artifact updates, to a dedicated runtime.
Its execution engine exploits GR's short token sequences, large candidate populations, and predictable step shapes to optimize self-attention, candidate selection, and execution planning.
\sys has been deployed for over nine months across our production lines, serving applications with hundreds of millions of users and supporting more than ten GR model families.
On our OneModel production workloads, \sys reaches over 1{,}000 QPS per serving instance with P99 latency below 35 ms and a 2.88$\times$ throughput improvement over representative baselines.
\end{abstract}

\settopmatter{printfolios=true,printacmref=false}
\maketitle
\pagestyle{plain}

\input{sec/01-introduction}

\input{sec/02-background}

\input{sec/03-design}
\input{sec/04-operator-specialization}
\input{sec/05-evaluation}
\input{sec/06-lessons}
\input{sec/07-related-work}
\input{sec/08-conclusion}

\clearpage
\bibliographystyle{ACM-Reference-Format}
\bibliography{references}

\end{document}

%% file: sec/01-introduction.tex
\section{Introduction}

Recommendation systems (RecSys) are core inference infrastructure for modern online services.
They shape what users see across discovery surfaces, from information feeds and short-video streams to search, e-commerce, and advertising placements~\cite{ying2018graph, pan2023learning,zhou2018deep, chen2025unisearch, chen2026onesearch}. Because these decisions directly affect user engagement and monetization, recommendation has grown into a large ecosystem of models and serving stacks~\cite{covington2016deep,cheng2016wide,zhang2019deep,yang2025gpu}.
This scale and diversity make recommendation serving a first-class systems problem, as shifts in RecSys model architecture quickly surface as changes in corresponding production infrastructure~\cite{gupta2020architectural}.

\begin{figure}[t]
  \centering
  \includegraphics[width=\columnwidth]{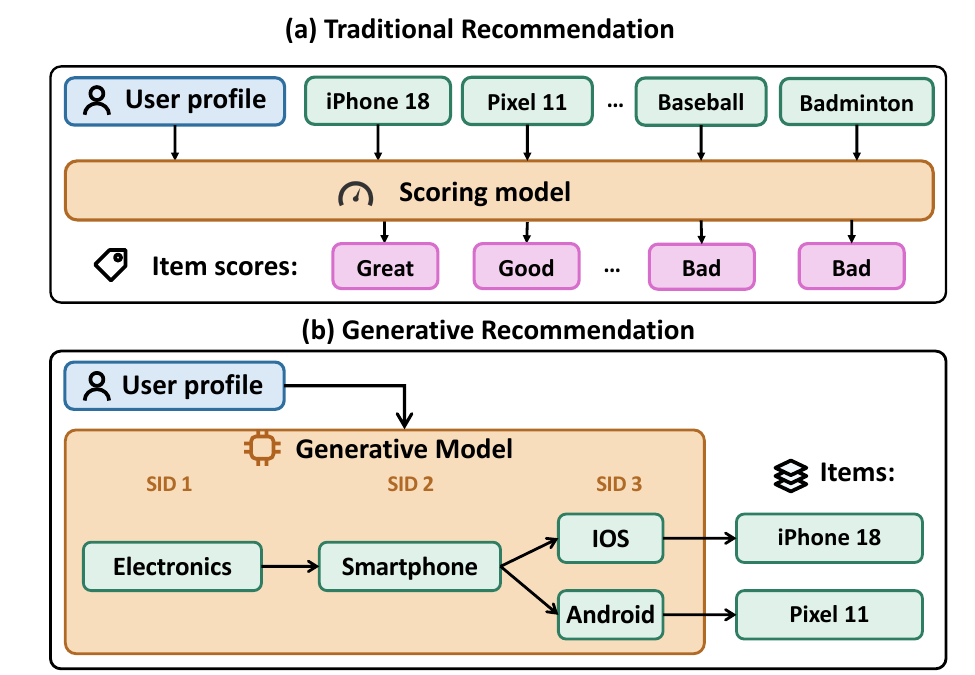}
  \vspace{-7mm}
  \caption{Traditional vs.\ generative recommendation.}
  \Description{A cascaded DLRM workflow that scores a shrinking candidate set, beside a generative workflow in which one model decodes the item's semantic identifier.}
  \label{fig:workflow}
  \vspace{-2mm}
\end{figure}

Serving recommendation workloads is challenging because each request must choose relevant items from an enormous catalog under interactive latency and high request volume~\cite{song2022rethinking,gupta2020architectural}.
As shown in Figure~\ref{fig:workflow}(a), traditional RecSys approach this task by scoring candidate items from the catalog according to their relevance to the user profile~\cite{cheng2016wide,naumov2019dlrm,song2022rethinking}.
This practice creates a long-standing systems compromise: broad item exploration must remain coarse and cheap to meet serving constraints, reducing the computation available to evaluate each item and, consequently, limiting the recommendation quality achievable at scale~\cite{song2022rethinking}.

The emerging \textit{generative recommendation} (GR)~\cite{zhou2025onerec,zhou2025onerecv2,zhai2024actions,han2025mtgr} paradigm challenges this conventional wisdom directly.
Rather than scoring supplied candidate items, GR autoregressively generates item identifiers from user context.
In the common formulation shown in Figure~\ref{fig:workflow}(b), each item is represented by a short \emph{semantic identifier} (SID) sequence~\cite{rajput2023tiger,singh2024semanticids}.
Generating SID tokens progressively narrows the item space toward specific items, allowing the model to guide item discovery directly~\cite{rajput2023tiger,deng2025onerec,zhou2025onerec,zhou2025onerecv2}.
This direction has gained industrial momentum with systems such as OneRec deployed in production~\cite{deng2025onerec,zhou2025onerecv2,han2025mtgr,agarwal2025pinrec,xue2026gr4ad,zhang2026onemall}.
GR therefore shifts the model serving challenge from scoring items in parallel to completing a chain of dependent generation steps within the same interactive latency budget.

\begin{figure}[t]
  \centering
  \includegraphics[width=0.9\columnwidth]{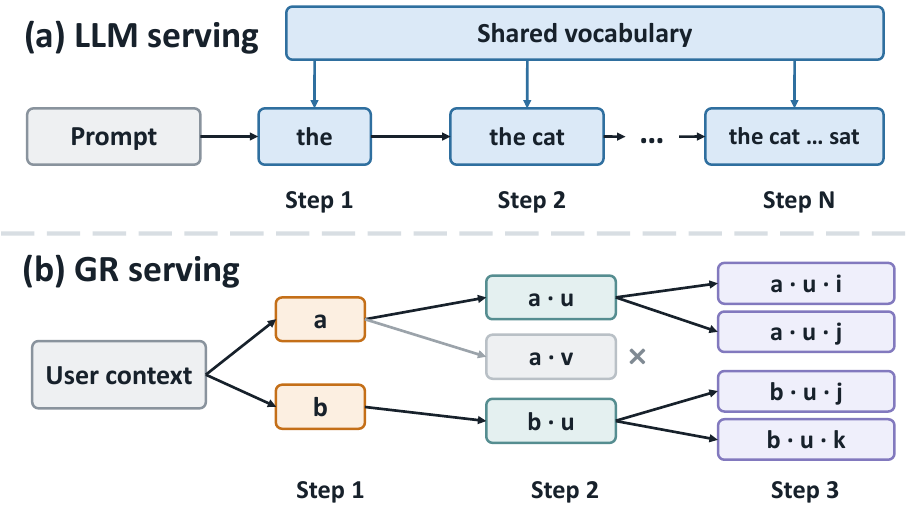}
  \vspace{-3mm}
  \caption{LLM serving vs.\ GR serving.}
  \Description{llm vs gr}
  \label{fig:llm-vs-gr}
  \vspace{-4mm}
\end{figure}

A natural question is whether GR serving can simply reuse the serving stack built for LLMs. After all, GR and LLMs both rely on autoregressive decoding, and recent LLM serving systems have made substantial progress in request scheduling, KV-cache management, and optimized decode execution~\cite{yu2022orca,kwon2023pagedattention,nvidia2026tensorrtllm,Wang2025KVCacheCI,Hu2025DEEPSERVESL,Gim2025PieAP}. 
As shown in Figure~\ref{fig:llm-vs-gr}(a), LLM serving systems benefit from a highly regular serving pattern: after the initial context is processed, each decoding iteration performs essentially the same task---consume the current sequence state, predict the next token from the same token space, append the result, and repeat. This regularity allows an LLM request to be handled behind a largely homogeneous decoding loop. GR appears to fit the same abstraction at first glance, while in fact it  violates the LLM serving assumptions in fundamental ways:

The first mismatch is in serving architecture.
As shown in Figure~\ref{fig:llm-vs-gr}(b), GR follows a short, fixed sequence of decoding steps that differ in both semantics and tensor shapes.
In terms of semantics, each step uses a distinct vocabulary, with earlier steps making coarse item distinctions and later ones refining them toward specific items.
In terms of shapes, each request must produce a set of recommended items, so a step needs to expand its input sequences into a larger set of candidate continuations.
The serving logic therefore needs to select which candidates survive based on predefined policies and prepare their associated state as inputs to the next step.
Frequent online updates then add a second pressure:
GR inherits the requirements of conventional RecSys~\cite{xie2020kraken}, and its model weights, per-step vocabularies, and selection policies evolve frequently under live traffic.
Together, heterogeneous step transitions and online update requirements strain existing serving frameworks built around homogeneous decoding loops and relatively stable artifacts.

The second mismatch is in execution optimization.
LLM serving stacks are typically optimized for workloads with long-context attention, largely uniform decode iterations, and small selection tails~\cite{kwon2023pagedattention,nvidia2026tensorrtllm,Wu2024LoongServeES}.
GR exposes a different set of execution demands.
In terms of attention computation, GR has many independent rows each attending over a short candidate sequence, leaving little work per row to amortize the overhead of attention kernels optimized for long sequences~\cite{dao2022flashattention}.
In terms of execution planning, different GR steps expose distinct but predictable shape profiles under a given deployment configuration, so a broad dynamic execution plan spanning all steps leaves step-level optimization opportunities unexplored.
In terms of candidate selection, expanding many sequences over per-step vocabularies produces a large pool of continuation scores from which multiple candidates are retained.
This makes selection computationally substantial on the critical path, rather than a lightweight sampling tail.
These three patterns limit the effectiveness of existing LLM execution strategies and motivate specialized optimizations for GR workloads.

To address these challenges, we present \textbf{\sys}, a production serving system for generative recommendation.
On the architecture side, \sys follows a \textit{runtime--engine decoupling} principle: a dedicated runtime coordinates decoding steps, manages candidate state, and handles online updates, while the engine focuses on tensor computation.
A deployment contract connects the two by specifying the per-step interfaces through which the runtime invokes the engine.
On the execution side, \sys specializes execution for GR's short prefixes, large candidate populations, and predictable step shapes.
\textit{Row-tiled self-attention} groups independent short attention rows to amortize execution overhead.
\textit{Step-granular planning} specializes execution plans to the narrower shape profiles of individual steps.
\textit{Fast candidate selection} reduces the cost of selecting survivors from the large set of candidate continuations.
The result is a GR serving stack that combines production-grade runtime operability with high throughput and low latency.

\sys has been deployed in production for over nine months, serving our applications with an average of 400 million daily active users across search, advertising, and recommendation scenarios. Over this period, \sys has supported more than ten GR model families, including OneModel, OneRec-v2, and UniSearch~\cite{zhou2025onerecv2, xue2026gr4ad, chen2025unisearch, chen2026onesearch}. It now also serves as the testing substrate for our upcoming GR models.
Evaluations on our OneModel production workloads demonstrate that \sys
reaches over 1{,}000 QPS per serving instance and P99 latency of under 35 ms, marking a 2.88$\times$ throughput improvement over representative baselines.
At cluster scale, \sys sustains a peak traffic of 60,000 QPS, supports high frequency online updates, and achieves high hardware efficiency with an MFU of 55.3\%. More broadly, the evolution of \sys shows the shift of performance bottlenecks as GR serving practices become mature, and a clear trajectory from adapting conventional RecSys and LLM serving stacks to building a GR-native system architecture.

%% file: sec/02-background.tex
\section{Background and Motivation}
\label{sec:background}
In this section, we motivate our problem further both in its real-world relevance and the choice of our techniques.

\subsection{The Rise of Generative Recommendation}
\label{subsec:background-autoregressive-gr}

Industrial recommendation has historically been served as a cascade of specialized scoring models, with earlier stages narrowing a very large item catalog and later stages scoring a smaller set of items with richer user context and item features~\cite{covington2016deep,song2022rethinking,naumov2019dlrm,gupta2020architectural}.
Recently, generative recommendation (GR) has emerged as a promising alternative to this conventional pipeline~\cite{he2026plum,zhai2024actions,zhou2025onerec}.
Unlike traditional scoring models that rank supplied candidate items, GR directly formulates recommendation as a generation problem: given user context, the model generates identifiers of items to recommend.
Early work explored generative and transducer-style recommendation~\cite{rajput2023tiger,zhai2024actions}, while the OneRec model family~\cite{deng2025onerec,zhou2025onerec,zhou2025onerecv2} brought GR into large-scale industrial practice.

\begin{figure}[t]
  \centering
  \includegraphics[width=\columnwidth]{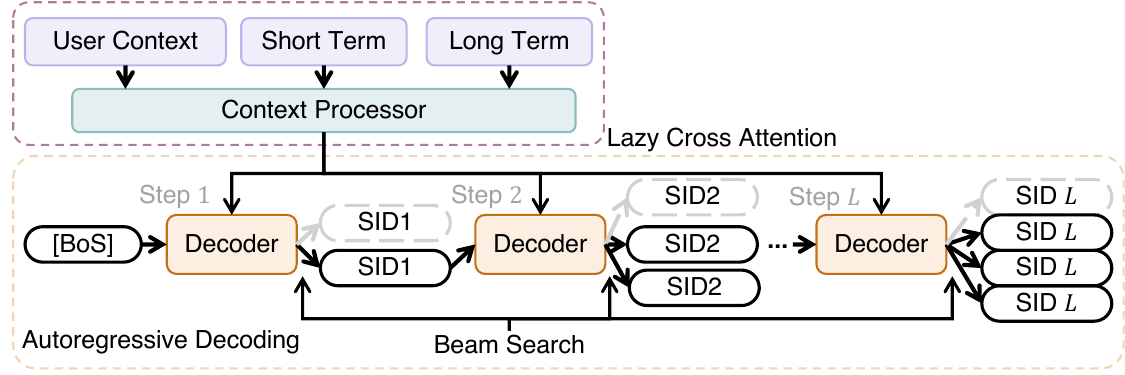}
  \vspace{-8mm}
  \caption{Execution of a GR request.}
  \Description{A two-part request diagram. Phase 1, context encoding, takes user static, short-term, and long-term pathways, passes them through a context processor, and produces a context representation or KV cache. Phase 2, autoregressive decoding, starts from BOS and repeatedly applies decoder blocks conditioned on the encoded context to generate SID tokens s1 through sL for one recommended item.}
  \label{fig:gr-request}
  \vspace{-5mm}
\end{figure}

As shown in Figure~\ref{fig:gr-request}, a GR request carries recommendation context, including static user information, short-term behaviors, and long-term interests.
The context processor encodes this information once into a representation reused as cross-attention state at every decoding step.
Each catalog item is represented by a short \emph{semantic identifier} (SID) sequence $(s_1,\dots,s_L)$, where each token $s_\ell$ is drawn from a position-specific vocabulary~\cite{rajput2023tiger,zhou2025onerec,zhou2025onerecv2,luo2025qarm}.
Decoding starts from a beginning-of-sequence (BoS) token and builds SID sequences one token at a time.
After each decoding step, candidate selection determines a limited set of the most likely candidate SID sequences and discards the rest.
This process is implemented through \textit{beam search}, which extends retained sequences with possible next tokens and selects the most likely resulting sequences.
After $L$ decoding steps, the top-$k$ completed SID sequences are mapped back to catalog items.

GR is attractive because it changes the modeling boundary of recommendation. By generating item identifiers directly, GR can optimize heterogeneous recommendation signals in one model and avoid losing target items in early coarse scoring stages~\cite{zhou2025onerec}. It also gives recommendation models a path to benefit from larger neural architectures and increased compute budgets; recent systems report strong quality gains and deployment momentum across major industrial scenarios, including applications from Kuaishou, Google, and Meta~\cite{deng2025onerec,he2026plum,zhai2024actions}. This paper focuses on the serving consequence of this shift: the production GR workload differs fundamentally from others.

\subsection{GR Requires a Specialized Serving Architecture}
\label{subsec:background-gr-neither-dlrm-nor-llm}

GR cannot directly reuse conventional RecSys or LLM serving stacks.
Conventional RecSys pipelines are highly specialized for models that take a set of items as input and return relevance scores.
Supporting autoregressive decoding therefore requires substantial redesign of their execution and state-management paths.
LLM serving offers a more natural starting point because it already supports autoregressive models with decoding loops.
The rest of this subsection explains why GR's step-transition and live-update requirements extend beyond the standard LLM decoding loop and motivate a GR-specific serving architecture.

\begin{figure}[t]
  \centering
  \includegraphics[width=\columnwidth]{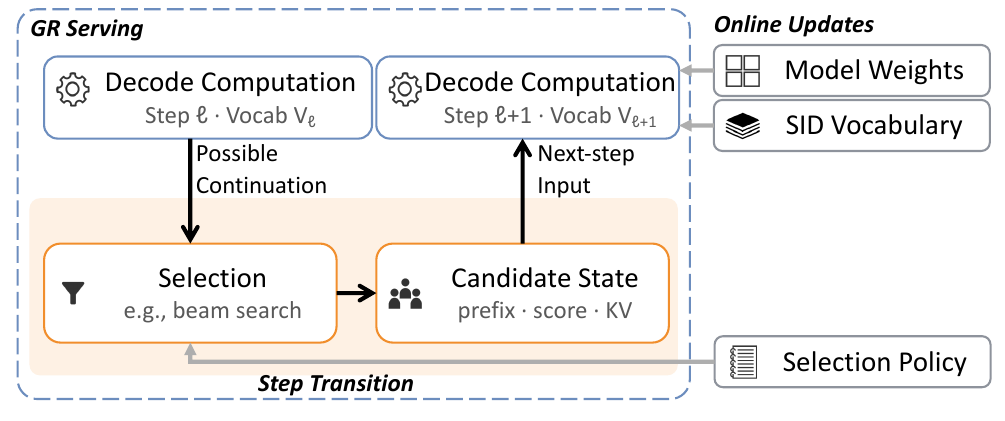}
  \vspace{-8mm}
  \caption{Heterogeneous transitions between decoding steps and online updates of serving artifacts.}
  \Description{Left: a GR serving box in which a compute engine runs decoding step l with vocabulary V_l and emits expanded candidates; a selection stage such as beam search keeps the surviving candidates; their candidate state (prefix, score, and KV) becomes the next-step input to the compute engine at step l+1 with vocabulary V_{l+1}. Selection and candidate state sit in a shaded step-transition region outside the engine. Right: online updates of model weights and the SID vocabulary point to the compute engine, and selection-policy updates point to the selection stage.}
  \label{fig:gr-serving-contract-pressures}
  \vspace{-4mm}
\end{figure}

\noindent\textbf{\textit{Observation 1: Heterogeneous step transitions.}}
To understand the architectural differences between LLM and GR serving, we compare the semantics and shapes of their individual decoding steps, as well as the transitions that connect these steps, as illustrated in Figure~\ref{fig:llm-vs-gr}.

\vspace{-1mm}
\setlist[itemize]{leftmargin=*}
\begin{packeditemize}

\item \textbf{Per-step semantics.}
In LLM decoding, each step predicts the next token from the same vocabulary.
In GR, each step predicts the next SID token from a step-specific vocabulary, with different positions carrying different levels of item meaning.
Earlier SID tokens distinguish coarse regions of the item space, while later tokens refine them toward specific items.
Intuitively, this resembles first choosing broad categories such as ``smartphone'' and then narrowing them down to products like ``iPhone 18''.

\item \textbf{Per-step shapes.}
While a standard LLM decoding iteration extends each active sequence with one token, a GR step may expand each input SID sequence into multiple candidate continuations.
Intuitively, a GR request may select several different products from the ``smartphone'' category.
The input and output sequence counts therefore differ within a step and vary across steps, as do the shapes of their associated state such as per-sequence KV cache.
These shapes are heterogeneous but predictable, as each step follows prescribed expansion logic.

\item \textbf{Step transitions.}
As shown on the left side of Figure~\ref{fig:gr-serving-contract-pressures}, the input to step $\ell{+}1$ is not simply the output of step $\ell$.
A selection policy, typically a beam-search variant, first decides which candidates survive.
The serving system then prepares each survivor's \emph{candidate state} for the next step: its updated SID prefix and accumulated score, together with the KV cache inherited from its parent sequence.
Intuitively, if a prefix representing ``smartphone'' has two surviving continuations representing ``iOS'' and ``Android'', both inherit the smartphone prefix's KV cache and extend it independently in subsequent decoding steps.
If no continuation survives, the parent prefix's cache is no longer needed and is excluded from subsequent decoding steps.
\end{packeditemize}

\noindent Together, these differences make GR serving a coordination problem: each step's model execution is connected to the next through policy-driven candidate selection and state preparation.
Simply wrapping an LLM engine with GR-specific transition logic relegates this central serving responsibility to ad-hoc glue outside its decoding abstraction.
Heterogeneous step transitions therefore demand explicit architectural support for coordinating step execution and candidate state.

\noindent\textbf{\textit{Observation 2: Frequent online updates.}}
LLM serving often treats the model, vocabulary, and decoding policy as stable choices during a worker’s lifetime~\cite{yu2022orca,kwon2023pagedattention}. GR serving inherits the freshness requirements of conventional RecSys: its model weights, SID vocabularies, and selection policies can all change while traffic is live. 
The right part of Figure~\ref{fig:gr-serving-contract-pressures} shows these components and where each update takes effect.

\begin{figure}[t]
  \centering
  \begin{minipage}[t]{0.48\columnwidth}
    \centering
    \includegraphics[width=\linewidth]{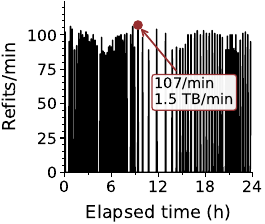}
  \end{minipage}
  \hfill
  \begin{minipage}[t]{0.48\columnwidth}
    \centering
    \includegraphics[width=\linewidth]{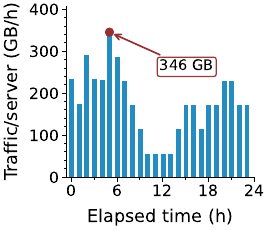}
  \end{minipage}
  \vspace{-3mm}
  \caption{Production online weight-refit activity in 24-hours: refit frequency (left) and FP32 refit traffic per server (right).}
  \Description{Two bar charts over a 24-hour capture window. The left chart shows model weight refits per minute, with a peak of 107 refits per minute. The right chart shows FP32 traffic per server in gigabytes per hour, with a peak of 346 gigabytes per hour.}
  \label{fig:online-weight-update-characterization}
  \vspace{-2mm}
\end{figure}

\vspace{-1mm}
\setlist[itemize]{leftmargin=*}
\begin{packeditemize}

\item \textbf{Model weights.}
Model weights cannot remain fixed for the lifetime of a serving worker in a RecSys deployment. Industrial recommendation has long required frequent dense-weight refresh~\cite{naumov2019dlrm,gupta2020architectural,yang2025prism, xie2020kraken}; we refer to each such online update of model weights on a live serving worker as a \emph{refit}. Figure~\ref{fig:online-weight-update-characterization} shows this pressure in our production GR cluster: online refits occur throughout the day, reaching 107 refits/min, while the corresponding FP32 refit traffic reaches 346~GB/hour per server in the busiest hourly bucket.
At this cadence, model weight refit cannot be handled as an offline redeployment procedure that drains traffic, replaces the serving engine, and then restores traffic.
The update must instead prepare the new weight version alongside the active pipeline, then publish it with minimal disruption to the request path.

\item \textbf{SID vocabulary.}
SID vocabularies encode item groupings learned from item representations.
As the item catalog evolves, the vocabularies need to be refreshed to reflect updated item relationships.
Such a refresh can change the meaning of existing tokens, which prefixes and continuations are valid, and which item each SID sequence identifies.
The serving pipeline must therefore keep each in-flight request's candidate SID sequences, per-step vocabularies, and final item mapping on a consistent version throughout decoding.

\item \textbf{Selection policy.}
GR exposes a broader and more volatile policy surface than ordinary LLM serving.
Apart from common policy space such as batch size and quantization precision, GR serving quality is also tied to the underlying candidate selection policy, which determines how candidates are expanded and retained across SID steps~\cite{xue2026gr4ad}. In our GR deployment, around twenty beam-search policy variants and accompanying hyperparameter settings have been exercised in production. Hard-coding each policy variant as part of a tightly coupled serving pipeline would tie policy iteration to complex and slow binary releases rather than flexible live updates.
\end{packeditemize}
\vspace{-1mm}

\noindent GR serving must roll out frequent updates to various artifacts with minimal disruption while keeping each request on a compatible set of artifact versions~\cite{bai2026pipelive,xie2020kraken}.
These updates create a second architectural pressure: artifact update and version consistency become an ongoing serving responsibility rather than a concern addressed only at deployment.

\vspace{-1.5mm}
\subsection{GR Exposes Distinct Optimization Opportunities}
\label{subsubsec:optimization-opportunities}

The discussion on serving architecture above exposes a second difference between GR and LLM serving: the optimization target changes~\cite{lai2023adaembed, du2026efficient}. LLM engines are tuned for long token streams, growing KV caches, and repeated decoding steps with similar execution shape. GR decode instead produces a short SID sequence over a large, beam-expanded set of candidates~\cite{zhou2025onerec, zhou2025onerecv2}. This shifts the dominant work away from long-sequence token generation and toward short, shape-sensitive stages whose costs depend on SID position, candidate population, and selection policy. The rest of this subsection explores the resulting opportunities for GR-specific performance optimizations.

\noindent\textbf{\textit{Observation 3: Attention dimension inversion.}}
GR uses the similar self-attention computation as LLM decode, but the dominant work shifts to different tensor dimensions. For a batch of sequences, self-attention reads a query tensor \(Q\) for the current positions and key/value tensors \(K,V\) for the positions that can be attended to. At a decoding step, this computation can be viewed as many independent attention rows: each live sequence, under each attention head, reads its own prefix. Using standard notation,
\begin{align*}
Q \in \mathbb{R}^{B\times S_q\times H\times D}, \quad
K,V \in \mathbb{R}^{B\times S_{\mathrm{kv}}\times H\times D},
\end{align*}
where \(B\) is the batch size, \(S_q\) is the number of current query positions, \(S_{\mathrm{kv}}\) is the attended-prefix length, and \(H\) and \(D\) are the number of heads and the per-head dimension.

\begin{figure}[t!]
  \centering
  \IfFileExists{figures/dimension_shape.pdf}{%
    \includegraphics[width=\columnwidth]{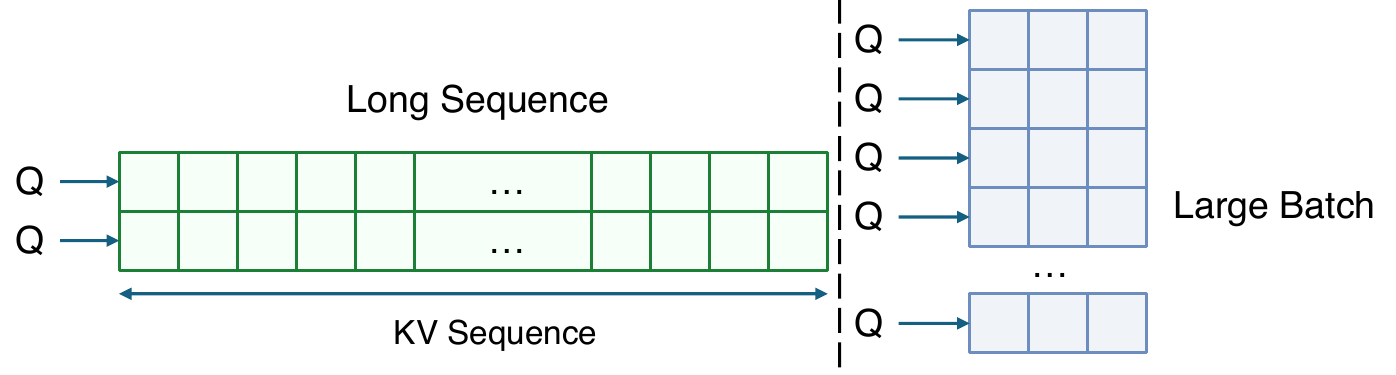}
  }{%
    \fbox{\begin{minipage}[c][1.35in][c]{0.92\columnwidth}
      \centering
      Placeholder for the dimension-inversion shape schematic.\\
      Missing: \texttt{figures/dimension\_shape.pdf}.
    \end{minipage}}
  }
  \vspace{-6mm}
  \caption{Attention dimension inversion in GR decode.}
  \Description{A two-panel schematic. The left panel shows LLM decode attention as a long horizontal KV sequence tile. The right panel shows GR decode as many short row-local attention rows and a row-local Top-K selection over a candidate width.}
  \label{fig:dimension-shape}
  \vspace{-4mm}
\end{figure}

In LLM serving, the main attention pressure often comes from \(S_{\mathrm{kv}}\): each live sequence may attend over a long prompt and generated history, giving kernels a long KV dimension to traverse and amortize over.
GR moves this pressure to the batch size \(B\).
Its attended prefix is only the short generated SID sequence, so \(S_{\mathrm{kv}}\) remains small.
Candidate selection via beam search instead multiplies the number of live sequences: a request batch \(B_{\mathrm{req}}\) with beam width \(W\) yields \(B=B_{\mathrm{req}}\times W\) candidate prefixes.
As shown in Figure~\ref{fig:dimension-shape}, GR attention has many more sequence rows, but each row attends over only a few SID positions.
This is the dimension inversion: LLM attention is commonly long-history attention, whereas GR attention is many-row short-history attention.

One might expect the large beam-expanded batch to make existing attention kernels efficient. The problem is that it creates many tiny, independent attention rows rather than one larger sequence tile.
Standard kernels~\cite{dao2022flashattention,ye2025flashinfer} expect long sequence tiles to amortize their setup costs.
GR exposes the opposite pattern: each row performs only a few dot products before a tiny softmax and value reduction.
The arithmetic in one row is too small to hide the overhead, yet these rows cannot be artificially merged into a single long sequence because each candidate requires strict isolation during softmax normalization.
The kernel opportunity is therefore structural: amortize overhead across many independent short rows while preserving each row's local reduction boundary.

\begin{figure}[t]
  \centering
  \includegraphics[width=\columnwidth]{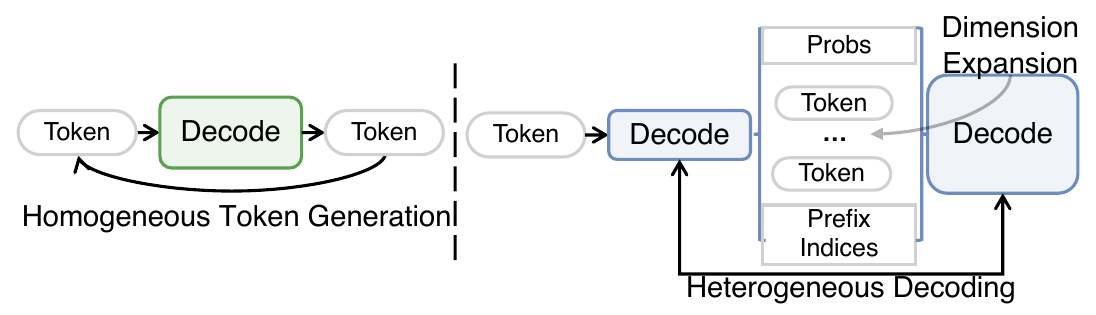}
  \vspace{-7mm}
  \caption{Decoding steps in LLM vs.\ GR serving.}
  \Description{A contrast between homogeneous token generation, which repeats the same decode transition from token to token, and heterogeneous GR decoding, where a decode step produces probabilities, tokens, and prefix indices and expands dimensions for the next step.}
  \label{fig:structured-decode-steps}
  \vspace{-4mm}
\end{figure}

\noindent\textbf{\textit{Observation 4: Predictable step shapes.}}
Observation~1 shows that GR decoding steps differ in shape; for execution planning, what matters is that these shapes are also predictable.
A GR request generates a short, fixed-length SID sequence, so the number of steps is known ahead of time~\cite{zhou2025onerec, han2025mtgr}.
Under a given deployment configuration, the prefix length, retained beam width, and candidate width of each step are also fixed in advance.
As illustrated in Figure~\ref{fig:structured-decode-steps}, early steps operate with short prefixes and limited candidate expansion, while later steps operate over beam-expanded prefixes whose size determines the shape of the next step.
Only traffic-dependent dimensions, such as the number of requests in a batch, vary at runtime.

LLM serving engines hide this predictability because they invoke the same token-level interface at every iteration~\cite{agrawal2024taming, yu2022orca}. Presented as one repeated operator, the few, fixed SID steps appear to have a wide range of possible shapes. The engine must then compile a single general plan that is valid for all SID positions, even though each position will execute in a much narrower, static profile. This is safe, but it leaves performance on the table: the workload has stable step-level structure, yet the engine sees only dynamic variation. The key opportunity is thus to keep these step boundaries visible as optimization boundaries, so compilation and scheduling can target the actual shapes of SID steps rather than one broad range that covers all steps.

\begin{figure}[t!]

  \includegraphics[width=0.7\columnwidth]{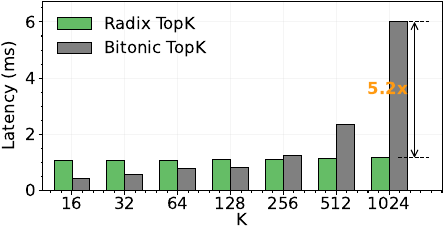}
  \vspace{-4mm}
  \caption{Candidate selection design choices.}
  \Description{A two-panel figure. Panel (a) shows Top-K dimension inversion: LLM decode selects a small \(K\), typically \(K<10\), from a wide token vocabulary, whereas GR beam update selects a large \(K\), typically \(K>128\), from a comparatively narrow semantic-id vocabulary on every step. Panel (b) shows a throughput comparison between radix-based Top-K and bitonic Top-K kernels across \(K\) values from 16 to 1024, illustrating that a single sampling-tail kernel choice does not dominate across the \(K\) range exercised by GR beam update.}
  \label{fig:topk-comparison}
  \vspace{-5mm}
\end{figure}

\noindent\textbf{\textit{Observation 5: On-path candidate selection.}}
In LLM serving, token selection is often a small tail, usually 1, after the model has produced logits~\cite{noarov2026foundations}.
In GR serving, however, selection sits squarely on the decoding critical path.
A SID step cannot finish by merely emitting scores; it must decide which SIDs survive to form the next step's input.
Because SIDs must explore a massive item space, GR beam search must retain a very large number of candidates at every step to preserve recommendation quality~\cite{zhou2025onerec}.
Consequently, this massive candidate expansion turns selection at each step transition into a computationally heavy, large-K operation that consumes a substantial fraction of serving time.

This large-\(K\) pattern makes GR especially sensitive to the choice of Top-K algorithm. A bitonic Top-K path performs structured compare-and-exchange ordering, which can be attractive for regular moderate-size selections but pays for ordering work that becomes costly as \(K\) grows~\cite{shanbhag2018efficient}. A radix Top-K path~\cite{li2024radik} instead narrows candidates by value ranges across passes, which can better match large-\(K\), many-row selection. Figure~\ref{fig:topk-comparison} shows this sensitivity under GR-style shapes: the choice between bitonic and radix selection can change latency by several times. This exposes a concrete Top-K optimization opportunity for GR serving: selection performance depends on building GPU-friendly selection paths that match the large-\(K\), many-row regime, rather than relying on a fixed LLM-style sampling implementation.

\begin{figure}[t]
  \centering
  \includegraphics[width=0.9\columnwidth]{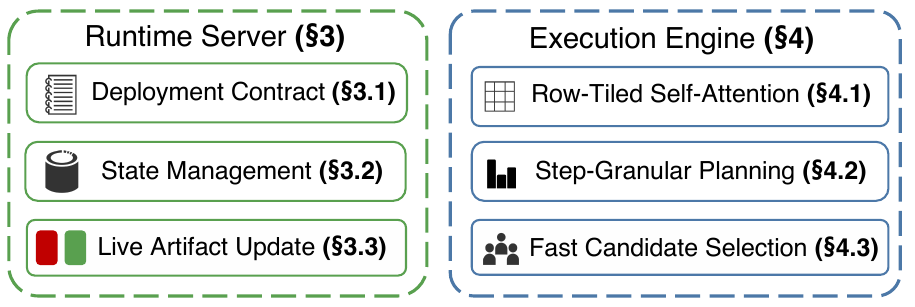}
  \vspace{-3mm}
  \caption{\sys serving system overview.}
  \Description{A simplified system architecture diagram. Dense and sparse embeddings and a configuration contract feed an infer-server binary. The binary orchestrates GPU-resident \sys engine execution, model weights, and custom kernels, with a loop arrow indicating iterative execution.}
  \label{fig:recogem-overview}
  \vspace{-5mm}
\end{figure}

\vspace{-1mm}
\subsection{\sys: A Production GR Serving System}
The observations above motivate the system organization shown in Figure~\ref{fig:recogem-overview}. To address the two architectural pressures in \textbf{Obs1–2}, \sys follows a runtime–engine decoupling principle. The runtime server owns the GR-specific serving responsibilities that a standard token loop does not express: it generates and follows a GR deployment contract, manages the transitions between SID steps, including candidate state, and applies online updates.
To exploit the performance opportunities in \textbf{Obs3–5}, \sys keeps SID step boundaries visible to the execution engine instead of collapsing generation into one homogeneous dynamic-shape loop. It then uses these boundaries to specialize the work inside a step, across steps, and after each step through row-tiled self-attention, step-granular planning, and fast candidate selection.

%% file: sec/03-design.tex
\section{\sys Serving Architecture}
\label{sec:serving-contract}

In this section, we outline the system architecture of \sys.
To handle heterogeneous step transitions (\textbf{Obs1}) and frequent online updates (\textbf{Obs2}), \sys keeps the execution engine narrow and moves GR-specific serving responsibilities into a dedicated runtime.
The engine executes tensor computation, while
the runtime manages the transitions between steps, including per-request candidate state, and applies online updates.
This runtime--engine decoupling explicitly preserves SID step boundaries, rather than collapsing the request into one opaque and generic token loop, which further enables \sys to exploit various performance optimization opportunities (\textbf{Obs3--5}).

\begin{figure}[t]
  \centering
  \includegraphics[width=01.00\columnwidth]{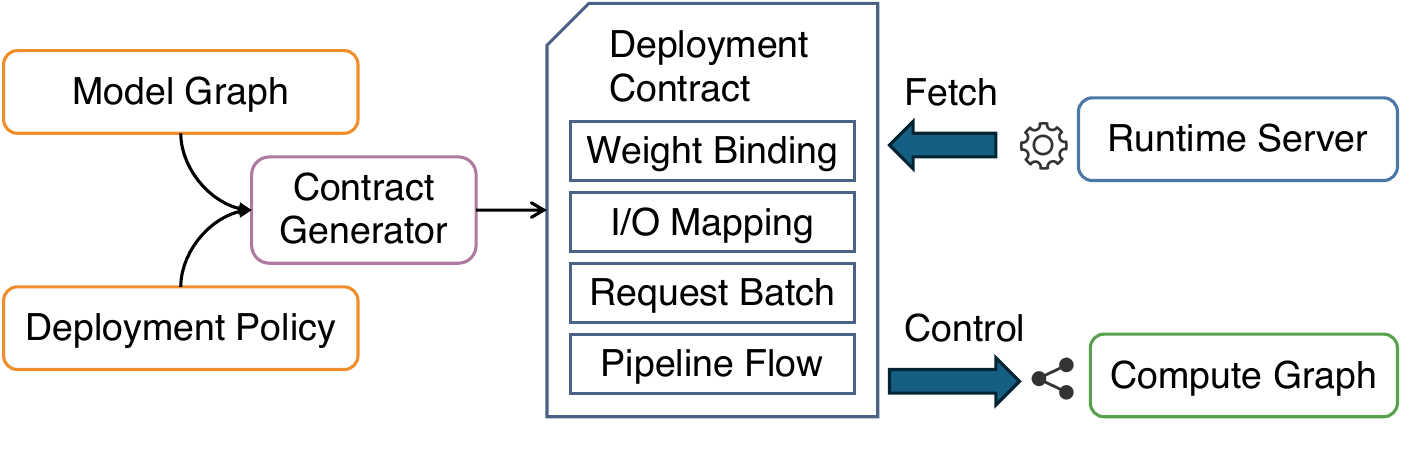}
  \vspace{-8mm}
  \caption{\sys deployment contract.}
  \Description{A diagram showing a model graph and deployment policy feeding a contract generator, which produces a deployment contract. The contract contains weight binding, input-output mapping, request batching, and pipeline flow records. The runtime server fetches the contract and uses it to control execution of the compute graph.}
  \label{fig:recogem-architecture}
  \vspace{-4mm}
\end{figure}

\vspace{-1mm}
\subsection{Deployment Contract Generation}
\label{sec:deployment-contract}
To make this decoupling work under frequent GR model updates, \sys needs a stable way to invoke and refit execution engines without reconstructing model-specific logic in runtime code.
\sys therefore automatically generates a deployment contract from the model graph and a small serving-policy overlay. The model graph is a good starting point because it exposes what the engine can legally consume, produce, and carry across serving I/O and SID steps. The policy overlay adds the deployment choices that are outside tensor computation, such as batch size, quantization, candidate size, and parallel layout. The generated contract combines these two views into the runtime-facing interface shown in Figure~\ref{fig:recogem-architecture}.
The runtime fetches the contract when loading a deployment and uses it to prepare the engine invocation path. It checks the contract against the compiled engine artifact on weight binding, I/O mapping, batch size, and pipeline flow, then uses the contract to guide request setup, step-to-step execution, and live updates.

\vspace{-1mm}
\subsection{Runtime State Management}
\label{sec:runtime-state-management}
\begin{figure}[t]
  \centering
  \includegraphics[width=\columnwidth]{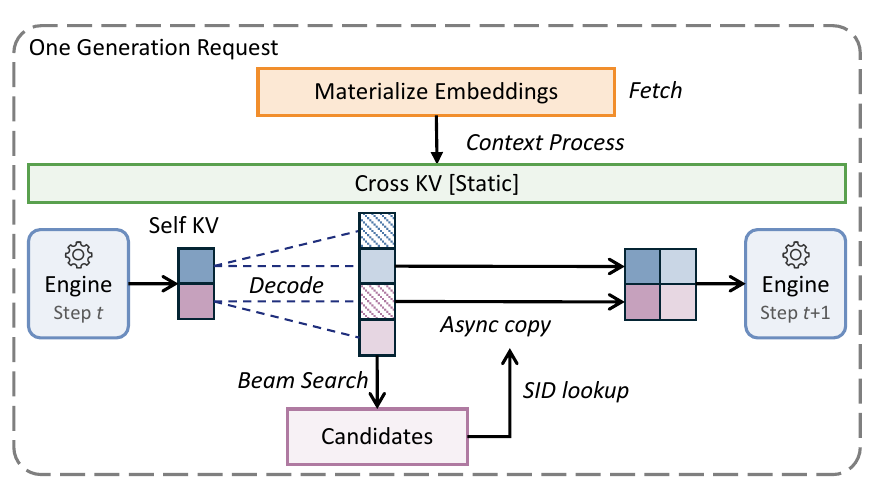}
  \vspace{-4mm}
  \caption{Runtime state management for a GR request.}
  \Description{A timeline-style diagram showing a \sys batch as contract-defined GPU tensor slices. Request inputs are copied into the batch buffer before enqueue. Context, self-KV, candidate ids, probabilities, and cache-indirection tensors are carried across decode steps by output-to-input mappings. Final outputs are copied to request buffers before the batch buffer is returned to the pool.}
  \label{fig:runtime-state-lifecycle}
  \vspace{-5mm}
\end{figure}

Figure~\ref{fig:runtime-state-lifecycle} shows how \sys manages the state of one GR request across its decoding steps.
The runtime server first materializes recommendation embeddings, and the engine converts the embeddings into cross-KV through context processing.
This cross-KV stays fixed for the whole request: once produced, it is reused by all SID decoding steps.
Self-KV follows a different lifecycle.
It is attached to candidate rows, and beam search changes which rows remain live after every step.
Candidate selection therefore creates sparse liveness in self-KV: some KV slices correspond to surviving candidates, while others belong to candidates that have been pruned.

\sys turns this sparse candidate state into a dense next-step KV prefix through double-buffered compaction.
At step \(t\), the engine decodes candidate continuations and invokes beam search to return the surviving candidates. The runtime
then prepares the next SID decoding step with a deterministic gather-and-copy pipeline.
It first uses the survivor indices to gather only the corresponding self-KV slices from the active KV buffer and copy them into a standby prefix buffer. This standby buffer then becomes the past-KV input for step \(t+1\).

In this way, the runtime realizes the contract's output-to-input mapping physically: candidate selection determines which KV slices remain live, while the next engine invocation sees a compact KV prefix rather than a sparse buffer that still holds pruned candidates.
It is worth noting that
the compaction loop is pipelined across survivor KV slices.
For each survivor, the runtime issues a copy into the standby prefix buffer.
While that copy is in flight, the runtime prepares the source address for the next survivor based on its SID lookup results.
Thus each iteration overlaps movement of the current KV slice with preparation for the next one.

\vspace{-1mm}
\subsection{Live Artifact Update}
\label{sec:checked-online-update}

GR updates occur at two different levels. Frequent updates refresh engine-side artifacts, such as model weights or SID vocabularies, while less frequent policy changes may also alter the deployment contract by changing the valid shapes, profiles, or step flow. \sys handles both cases through the same rule: a new serving version becomes visible only after it passes the contract gate. For weight or vocabulary refreshes, the runtime validates the new engine artifact against the active contract; for policy changes, it validates the new engine against the corresponding new contract.

The runtime prepares the next serving version outside the active request path. During cutover, it briefly stops admitting new GR requests, lets already-issued requests finish on the old version, and then atomically switches the active engine pointer, along with the contract pointer when the contract changes. Queued requests are then released to the new version, and the old version is retired after it is no longer referenced. This mechanism is intentionally simple: the important property is that each request executes under one consistent model/vocabulary/policy view, without coarse traffic draining or server-binary changes.
Together, the generated contract, state management, and live updates define the runtime boundary around the tensor engine.

%% file: sec/04-operator-specialization.tex
\section{\sys Performance Optimization}
\label{sec:engine}
\label{sec:optimization}
\label{sec:step-execution}
\label{sec:kernel-specialization}
\sys addresses the three performance mismatches in GR serving.
Within each decoding step, dimension inversion turns self-attention into many short, independent row reductions rather than a long-sequence operator (\textbf{Obs3}). Across SID steps, generation follows a small, fixed number of heterogeneous regimes, so treating all steps as one dynamic case discards useful structure (\textbf{Obs4}). After each SID step, candidate selection determines which candidates enter the next step, making it part of the critical path rather than a lightweight sampling tail (\textbf{Obs5}). \sys therefore specializes from these three complementary angles.

\subsection{Row-Tiled Self-Attention}
\label{subsec:row-tiled-short-sequence-attention}

\begin{figure}[t]
  \centering
  \includegraphics[width=\columnwidth]{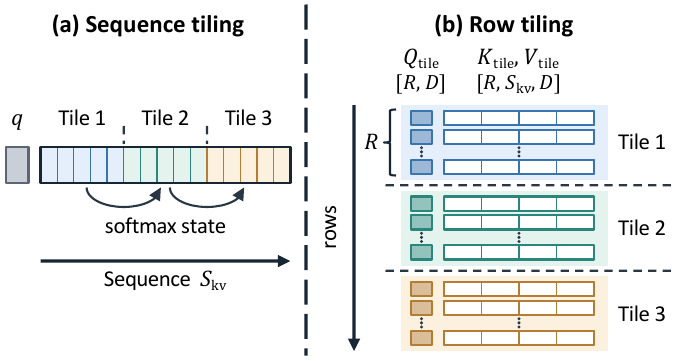}
  \vspace{-6mm}
  \caption{Sequence tiling versus row tiling.}
  \Description{Sequence tiling splits one long attention row; row tiling groups R complete short rows per tile.}
  \label{fig:attention-tiling}
  \vspace{-4mm}
\end{figure}

GR self-attention exposes many independent rows, each with a short KV prefix.
We fold the beam-expanded batch and head dimensions, so each row contains one query $q\in\mathbb{R}^{D}$ and its cached keys and values $k,v\in\mathbb{R}^{S_{\mathrm{kv}}\times D}$.
Figure~\ref{fig:attention-tiling} contrasts the tiling strategies: sequence tiling splits a long row along its KV sequence~\cite{dao2022flashattention,ye2025flashinfer}, whereas \sys groups $R$ complete short rows into one execution tile.
This grouping amortizes setup overhead across independent rows.
Each tile stacks the queries, keys, values, and outputs of its $R$ rows into
\begin{align*}
Q_{\mathrm{tile}}\in\mathbb{R}^{R\times D},\quad
K_{\mathrm{tile}},V_{\mathrm{tile}}\in\mathbb{R}^{R\times S_{\mathrm{kv}}\times D},\quad
O_{\mathrm{tile}}\in\mathbb{R}^{R\times D}.
\end{align*}

\begin{figure}[t]
  \centering
  \includegraphics[width=0.95\columnwidth]{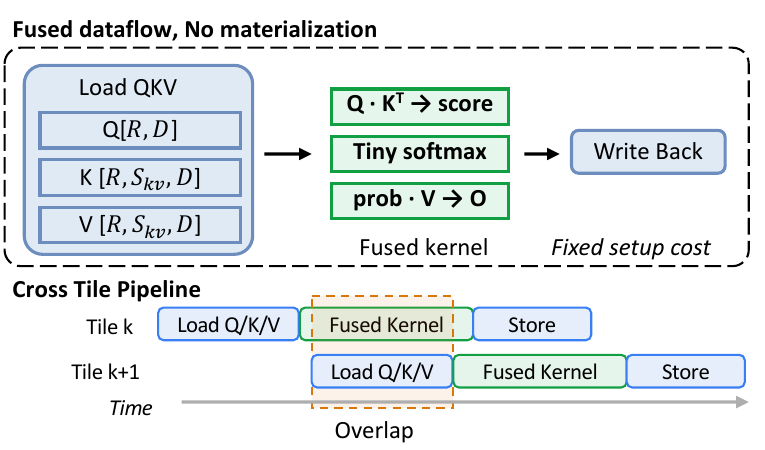}
  \vspace{-4mm}
  \caption{Fused attention dataflow and cross-tile pipelining.}
  \Description{Top: fused per-row attention keeps intermediates on chip. Bottom: loading tile k+1 overlaps computation on tile k.}
  \label{fig:short-sequence-attention}
  \vspace{-6mm}
\end{figure}

Figure~\ref{fig:short-sequence-attention} shows how these tiles execute.
Because $S_{\mathrm{kv}}$ is small, each row's entire attention reduction can be completed on chip.
Within each tile, the kernel stages $Q_{\mathrm{tile}}$, $K_{\mathrm{tile}}$, and $V_{\mathrm{tile}}$ into on-chip memory and fuses score computation, softmax over $S_{\mathrm{kv}}$ positions, and value reduction for each row.
Intermediate scores and probabilities remain on chip; only $O_{\mathrm{tile}}$ is written back.

Sequence tiling must maintain softmax normalization state across tiles belonging to the same long row.
In contrast, each short row's softmax is complete within its tile, so tiles need no normalization state from one another.
The kernel therefore overlaps the $Q/K/V$ load of tile $k+1$ with the fused computation of tile $k$, as shown in the lower panel.
This pipeline reduces exposed memory-staging latency while preserving independent per-row attention.

\subsection{Step-Granular Planning}
\label{subsec:step-granular-execution-planning}

\begin{figure}[t]
  \centering
  \includegraphics[width=\columnwidth]{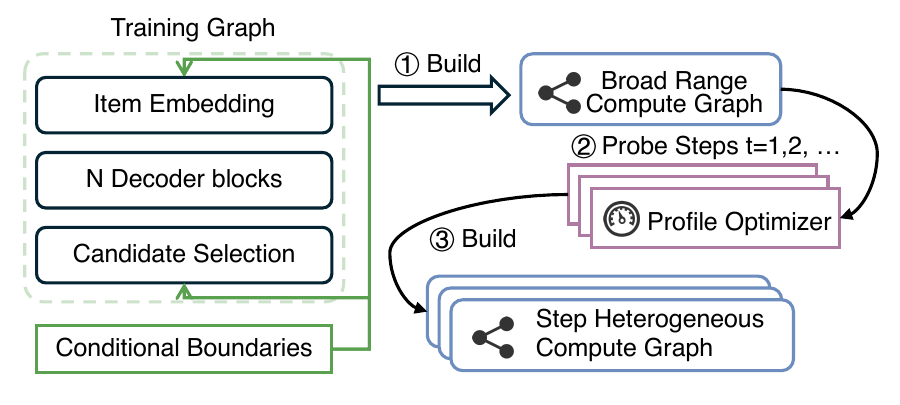}
  \vspace{-2mm}
  \caption{Step-granular execution planning lowers a single-step GR graph into step-specialized engines by probing per-step shapes and rebuilding under step-bounded profiles.}
  \Description{A left-to-right planning pipeline. On the left, a dashed box labeled Training Graph stacks Item Embedding, N decoder blocks, and Candidate Selection, with a Conditional Boundaries box feeding back into the graph. The first Build stage turns this graph into a Broad Range Compute Graph. The Probe Steps stage, over generation steps t equals 1, 2, and onward, passes it through a Profile Optimizer that records per-step shapes. The second Build stage emits a stacked Step Heterogeneous Compute Graph specialized for each generation step.}
  \label{fig:compile-pipeline}
  \vspace{-2mm}
\end{figure}

GR decoding executes a short sequence of SID steps with predictable but different tensor dimensions.
A single broad dynamic profile can cover all steps, but it forces the engine to optimize for the union of dimensions that appear across all SID positions.
\sys instead turns these recurring step differences into build-time profile boundaries.
The goal is to keep one logical GR step as the model interface, while producing step-specialized execution artifacts for the shapes that actually occur during generation.
Figure~\ref{fig:compile-pipeline} shows the three-stage planning pipeline.

First, \sys builds a broad-range compute graph from the training graph and its conditional boundaries.
This intermediate graph is intentionally permissive: it can represent the tensor dimensions that may appear at any SID step, but it is not the final optimized artifact.
Second, \sys performs a compile-time dry run over logical steps $t=1,2,\ldots,L$.
At each step, the planner follows the graph’s conditional boundaries and output-to-input tensor links, infers the dimensions of the tensors produced by that step, and feeds those dimensions into the next step.
Third, the profile optimizer converts the collected per-step dimension table into step-bounded profiles and generates the final step heterogeneous compute graph.
The final graph is therefore optimized for the narrower dimensions each SID step actually presents, rather than for one broad dynamic range spanning all steps.
It is worth noting that traffic-induced dimensions such as batch size remain dynamic, so the compiled artifact can still serve variable traffic patterns.

To see how step-granular planning improves engine execution, consider the row-tiled attention path inside each SID decoding step.
During the compile-time dry run, \sys records the cached-prefix length \(S_{\mathrm{kv}}\) that each step presents to the self-attention.
The engine can then bind that step to a distinct row-tiled attention variant whose tiny-softmax length and \(K/V\) staging extent match the recorded \(S_{\mathrm{kv}}\). Such specialization avoids a broad dynamic inner loop on the attention critical path.
As another example, the same dry run records candidate width and retained beam width after each decode step, exposing the concrete beam-search shapes that candidate selection can specialize on.
Thus step-granular planning is central to \sys's performance design: it turns predictable SID-step evolution into the narrow engine constraints consumed by various GR-specific operators.

\subsection{Fast Candidate Selection}
\label{subsec:large-candidate-topk-selection}

Candidate selection is on the critical path of every SID step: the next decode step cannot begin until beam search chooses the surviving partial SIDs.
In production GR, each surviving candidate is expanded over a large SID vocabulary, creating a large beam-search space for Top-K selection rather than a lightweight sampling tail.
The main question is therefore how to execute Top-K efficiently at every step boundary.
A strawman approach is to flatten all candidate scores of a request and run one large Top-K.
For one GR request, the scores form a $W\times C$ matrix: $W$ beams, each expanded to $C$ candidate continuations.
Flattening this matrix creates a single row of $W C$ scores and selects the next $K$ beams from it.
This is correct, but it gives the GPU the wrong physical shape, turning
candidate selection into a repeated global memory access problem.
As a concrete example, in one of our GR serving settings, $C=8192$ and $W=512$, so one flattened request contains \(4.2\) million candidates. Even with FP16, this request alone occupies \(8\) MiB, far beyond the local memory available to a GPU thread block.

\sys instead preserves the beam matrix boundary and performs the update in two exact stages.
First, it applies a per-row precut: from each beam row of width $C$, it keeps the local top $K$.
Second, it performs the request-level selection over the union of precut survivors to choose the next $K$ beams.
This preserves the exact global top-$K$ result: any candidate discarded by the per-row precut is already dominated by $K$ candidates from the same row, so it cannot appear among the final $K$ winners across the request.
Note that the per-row boundary is a natural cut for GR, because it can be reduced locally before the request-level merge, without introducing additional cross-row regrouping steps.

The two stages expose different GPU shapes. The per-row precut has many independent rows of bounded width, so \sys uses a flash radix path that treats each beam row as a local selection unit: row scores are staged into shared memory, radix passes run over the local row, and a group of multiple rows can be packed into one thread block.
The implementation also pipelines different row groups, loading the next group while searching over the current one. This turns the dominant first stage into efficient local selections.
After the precut, the request-level selection is still a standard Top-K, but its input has already been reduced from the original $W\times C$ candidates to the $W\times K$ row survivors. \sys uses a multi-block radix path that partitions one request-level row across blocks and merges their partial selections. Meanwhile,
step-granular planning supplies the per-step beam size, allowing the engine to specialize both stages to the paths that match their concrete beam-search shapes.

%% file: sec/05-evaluation.tex
\section{Evaluation}
\label{sec:eval}

\noindent\textbf{Implementation.}
We implement \sys in approximately 20K lines of C++/CUDA code and 30K lines of Python. \sys has been deployed for over nine months, and serves over 400 million daily active users. It has hosted more than ten GR model families, including both OneRec-v1 and OneRec-v2.
In the \sys deployment pipeline, we construct TensorRT engines from the PyTorch model graph, as the latter is the standard interface provided by the algorithm team.

\noindent\textbf{Evaluation questions.}
We organize the evaluation around four main questions:
(i) How much end-to-end performance gain does \sys achieve across production GR models?
(ii) How do \sys's design choices contribute to performance across heterogeneous SID steps?
(iii) How effectively does row-tiled attention exploit GR's short-prefix, beam-expanded regime?
(iv) To what extent does fast candidate selection reduce the overhead of large-\(K\) beam updates?
Additionally, we validate the performance of \sys under real-world traffic conditions.

\subsection{Experimental Setup}
\label{subsec:eval-setup}

Our experiments evaluate two production GR models:
\emph{OneRec-v2} is the higher-capacity deployment with 4B parameters, while \emph{OneModel} is the lighter deployment with 0.2B parameters. Both models share the same core architecture with minor structural variations.
These experiments are conducted on a host equipped with data-center GPUs and high-end Intel CPUs. The software stack includes Linux 5.14, CUDA 12.8, and TensorRT 10.16.0.72.
To demonstrate the benefits of our design, we compare four execution configurations:
\vspace{-1mm}
\setlist[itemize]{leftmargin=*}
\begin{packeditemize}
    \item \emph{\sys}: Our full system with all optimizations enabled.
    \item \emph{GeneralGen}: Uses a homogeneous shape profile and naive GPU kernels, while keeping other modules unchanged.
    \item \emph{TorchEager}: Executes the GR model in PyTorch eager mode.
    \item \emph{TRT-LLM}: Adapts TensorRT-LLM for GR serving.
\end{packeditemize}

\subsection{End-to-End Performance on GPUs}
\label{subsec:eval-e1-onerecv2}

\begin{figure}[t!]
  \centering
  \includegraphics[width=0.9\linewidth]{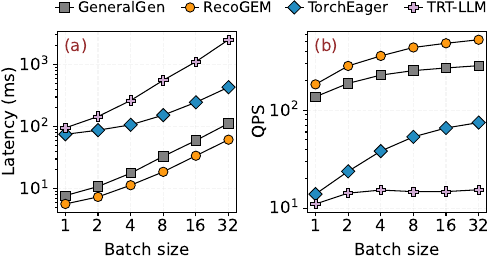}
  \vspace{-4mm}
  \caption{Performance trend of OneRec-v2.}
  \Description{Four curves over batch sizes 1 to 32 compare GeneralGen, \sys, TorchEager, and the adapted TensorRT-LLM runtime. The panels show P99 latency and QPS on logarithmic vertical axes.}
  \label{fig:eval-e1-onerecv2}
  \vspace{-4mm}
\end{figure}

\begin{figure}[t!]
  \centering
      \includegraphics[width=0.9\linewidth]{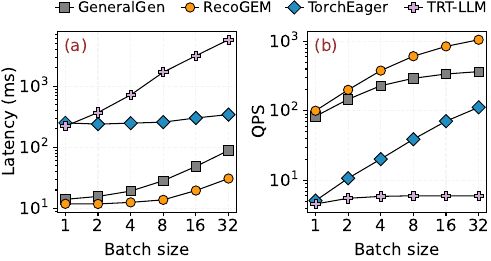}
  \vspace{-4mm}
  \caption{Performance trend of OneModel.}
  \Description{Four curves over batch sizes 1 to 32 compare GeneralGen, \sys, TorchEager, and the adapted TensorRT-LLM runtime with dynamic beam widths 128, 256, 512, and 1024. The panels show P99 latency and QPS on logarithmic vertical axes.}
  \label{fig:eval-e1-onemodel}
  \vspace{-4mm}
\end{figure}

To demonstrate the performance of \sys, we evaluate it with \emph{OneRec-v2} and \emph{OneModel} under various request batch sizes. %
Because recommendation is strictly constrained by tail latency, we focus on reporting the 99th percentile (P99) latency alongside the serving queries per second (QPS).

Figure~\ref{fig:eval-e1-onerecv2} shows the performance of \emph{OneRec-v2}, where \sys keeps tail latency strictly below the production 100\,ms target across the entire batch sweep.
At batch 32, it achieves a 61.5\,ms P99 latency and 526 QPS.
Under \emph{GeneralGen}, P99 latency rises to 112.5\,ms and QPS drops to 286, demonstrating the combined benefit of step-granular planning and operator specialization at large batch sizes.
Figure~\ref{fig:eval-e1-onemodel} shows the performance of \emph{OneModel}, where \sys achieves even lower latency and higher QPS due to the smaller model size.
At batch 32, \sys reaches a 30.8\,ms P99 latency and 1{,}053 QPS, compared to 89.0\,ms and 366 QPS for \emph{GeneralGen}.
The relative throughput gain over \emph{GeneralGen} is larger on \emph{OneModel}, reaching 2.88$\times$ compared to 1.84$\times$ on \emph{OneRec-v2}.

TorchEager and TensorRT-LLM exhibit substantially higher latency and lower throughput than \sys across both models.
TorchEager's gap is consistent with the dispatch, kernel-launch, and intermediate-tensor overheads of eager execution, which are costly for short SID sequences.
For the adapted TensorRT-LLM runtime, latency grows sharply with batch size while QPS quickly plateaus, indicating limited batching gains under the evaluated configuration.
Unlike TorchEager, which keeps all computations on the GPU, the TensorRT-LLM runtime performs beam search on the CPU and must then reorder the KV cache according to the selected beams, which adds heavy framework overhead at every SID step whose cost grows with the number of live beams.
These observations confirm the mismatch between LLM-oriented serving engines and GR's short-prefix attention, large candidate sets, and heterogeneous SID steps.

\begin{figure}[t]
  \centering
  \includegraphics[width=0.9\linewidth]{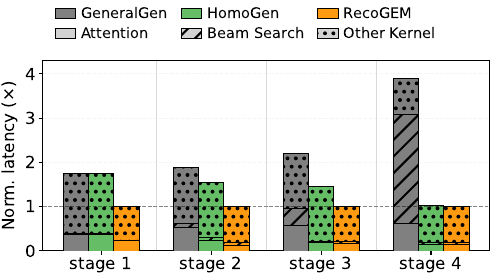}
  \vspace{-3mm}
  \caption{Batch-32 step-wise kernel breakdown for three serving systems in OneModel.}
  \Description{Four groups of stacked bars compare GeneralGen, HomoGen, and \sys across OneModel's four SID steps at batch size 32. Each bar separates attention, beam search, and other kernels. Each component is divided by \sys's total kernel latency at the same step, so every \sys bar totals one.}
  \label{fig:eval-e2-bs32-kernel-breakdown}
  \vspace{-4mm}
\end{figure}

\begin{figure*}[t]
  \centering
  \captionsetup[subfigure]{font=small,skip=2pt}
  \begin{subfigure}[t]{0.32\textwidth}
    \centering
    \includegraphics[width=\linewidth]{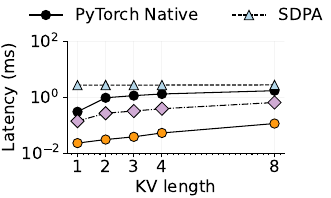}
    \vspace{-4mm}
    \caption{$B{=}4096$, $H{=}32$}
    \vspace{-2mm}
    \label{fig:eval-attention-b4096-h32}
  \end{subfigure}\hfill
  \begin{subfigure}[t]{0.31\textwidth}
    \centering
    \includegraphics[width=\linewidth]{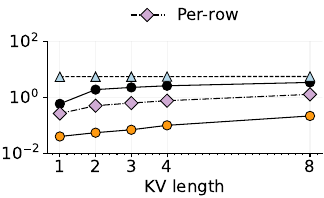}
    \vspace{-4mm}
    \caption{$B{=}8192$, $H{=}32$}
    \vspace{-2mm}
    \label{fig:eval-attention-b8192-h32}
  \end{subfigure}\hfill
  \begin{subfigure}[t]{0.31\textwidth}
    \centering
    \includegraphics[width=\linewidth]{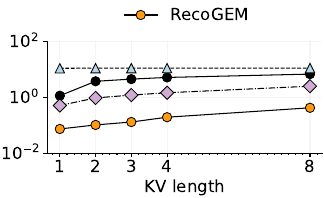}
    \vspace{-4mm}
    \caption{$B{=}16384$, $H{=}32$}
    \vspace{-2mm}
    \label{fig:eval-attention-b16384-h32}
  \end{subfigure}
  \par\vspace{0.6em}
  \begin{subfigure}[t]{0.32\textwidth}
    \centering
    \includegraphics[width=\linewidth]{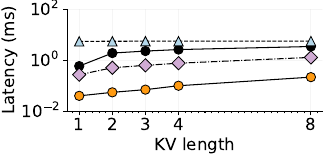}
    \vspace{-4mm}
    \caption{$B{=}4096$, $H{=}64$}
    \label{fig:eval-attention-b4096-h64}
  \end{subfigure}\hfill
  \begin{subfigure}[t]{0.32\textwidth}
    \centering
    \includegraphics[width=\linewidth]{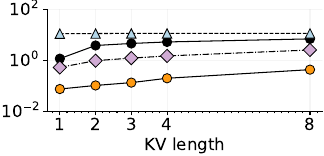}
    \vspace{-4mm}
    \caption{$B{=}8192$, $H{=}64$}
    \label{fig:eval-attention-b8192-h64}
  \end{subfigure}\hfill
  \begin{subfigure}[t]{0.32\textwidth}
    \centering
    \includegraphics[width=\linewidth]{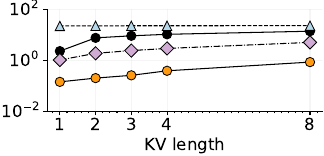}
    \vspace{-4mm}
    \caption{$B{=}16384$, $H{=}64$}
    \label{fig:eval-attention-b16384-h64}
  \end{subfigure}
  \vspace{-3mm}
  \caption{Decode self-attention latency for the GR short-prefix setting across batch and head sweeps.}
  \Description{Six log-scale line plots arranged in a two-by-three grid, comparing PyTorch Native, SDPA, the earlier per-row \sys path, and the current \sys row-tiled path across short key/value lengths. The top row fixes the number of heads at 32 and sweeps effective batch over 4096, 8192, and 16384; the bottom row repeats the sweep at 64 heads. \sys is the fastest backend in every panel.}
  \vspace{-3mm}
  \label{fig:eval-attention}
\end{figure*}

\subsection{Fine-Grained Performance Breakdown}
\label{subsec:eval-step-wise-breakdown}

Apart from end-to-end results, we further decompose serving execution time to precisely attribute \sys's performance gains to its major design choices.
Specifically, we evaluate the batch-32 operating point of \emph{OneModel} and categorize the end-to-end decode latency into attention, beam search (candidate selection), and all remaining GPU kernels (e.g., dense projections and MoE/FFN computation), as depicted in Figure~\ref{fig:eval-e2-bs32-kernel-breakdown}.
We evaluate \sys, \emph{HomoGen}, and \emph{GeneralGen} throughout this breakdown and measure per-step latency using the NVIDIA Nsight Systems (\texttt{nsys}) profiler.
Here, \emph{HomoGen} removes step-granular planning by using a single shape profile across all SID steps while retaining \sys's specialized kernels.

\emph{(i) The benefit of step-granular planning mainly comes from earlier decoding steps.}
\sys bounds step-derived dimensions such as past-key length, retained beam width, and candidate width separately for each step.
At batch 32, the resulting profile specialization reduces average decode latency from 39.5\,ms to 30.4\,ms for \emph{OneModel}.
These steps exercise distinct SID shapes, moving from the initial step to beam-expanded steps with different retained-beam and candidate widths; under \emph{HomoGen}'s single profile, this mismatch translates into over 40\% higher latency in the first three SID steps.

\emph{(ii) The benefit of optimized attention and beam-search kernels mainly comes from later decoding steps.}
\emph{GeneralGen} uses general attention and selection paths that are optimized for different regimes: sequence-oriented attention assumes enough query/key length to amortize a long-sequence tile, while generic Top-K paths do not match the two-stage, large-\(K\) beam update induced by SID generation.
\sys instead maps attention to many independent short-prefix rows and maps beam search to a radix-based per-beam precut followed by request-level global selection.
Consequently, the \emph{GeneralGen} bars grow primarily in the attention and beam search components of the later decode steps.
For \emph{OneModel}, the final step rises from 13.8\,ms under \sys to 54.0\,ms under \emph{GeneralGen}, dominated by candidate selection.
This breakdown explains why the end-to-end gap is not a generic tuning artifact: \sys gains by exposing the GR step shape and then implementing operators that best match that shape.

\subsection{Row-Tiled Attention Kernel Evaluation}
\label{subsec:eval-attention}

Based on the step-wise breakdown above, we further isolate the self-attention operator to evaluate the attention-side performance contribution to \sys.
We sweep GR attention settings to show how the row-tiled path performs across the intended regime: many independent batch/head rows with a short key/value prefix.
It is worth noting that \(B\) is the request batch size multiplied by the beam width, and \(H\) and \(D\) are the number of attention heads and the per-head dimension.
We compare four implementations in the same benchmark harness: an explicit PyTorch implementation reference (\emph{PyTorch Native}), PyTorch SDPA as the general attention path selected in this environment (\emph{SDPA}), an earlier \sys implementation that processes each \((b,h)\) row independently (\emph{Per-row}), and the current row-tiled kernel (\emph{\sys}).

Across Figure~\ref{fig:eval-attention}, \sys's row-tiled attention path is fastest in all six evaluated cases.
Figure~\ref{fig:eval-attention}(f) is the canonical high-beam GR decode-attention case in this sweep: \(B{=}16384\), \(H{=}64\), and \(D{=}64\), with one attention computation per row and a bounded SID-prefix length.
In this setting, \sys is consistently fastest across the tested short-prefix region, reaching 0.142\,ms at \(S_{\mathrm{kv}}{=}1\) and remaining below 0.85\,ms through \(S_{\mathrm{kv}}{=}8\).
Within the main prefix range \(S_{\mathrm{kv}}\le4\), the row-tiled path improves over the earlier per-row path by 7.2--9.4\(\times\), with a maximum 154.4\(\times\) speedup over the SDPA path observed in this harness.
The speedup reflects three structural effects:

\emph{(i) SDPA is mismatched to the short-prefix regime.}
Fused sequence-attention paths are designed to amortize their setup cost over long tiles.
When \(S_q{=}1\) and \(S_{\mathrm{kv}}\le8\), the useful work in each row is only a short dot-product, a tiny softmax, and one value reduction.
The long-sequence machinery still executes a much larger padded tile and pays boundary-handling and running-softmax overheads, so most of the work is unrelated to the few valid prefix positions.
This is also reflected in Figure~\ref{fig:eval-attention}: SDPA latency changes little as \(S_{\mathrm{kv}}\) grows within the short-prefix range, indicating that fixed tile overhead dominates the actual attention work.

\emph{(ii) The per-row path removes framework-level overheads.}
The per-row kernel fuses QK, softmax, and PV for one \((b,h)\) row, keeping scores and probabilities in registers instead of materializing intermediate tensors.
Because \(S_{\mathrm{kv}}\) is known at compile time, the softmax becomes a fixed short reduction rather than a generic masked operator.
This explains why the per-row path is much faster than both PyTorch baselines.

\emph{(iii) The row-tiled path amortizes row-level fixed cost.}
The per-row path still maps each \((b,h)\) row to an independent work unit; at the largest setting this creates roughly one million row instances.
Row-tiled attention instead packs 64 rows into each execution tile and keeps blocks alive across tiles with a persistent scheduler, amortizing block scheduling and pipeline setup across many row-local reductions.
Asynchronous staging further overlaps tile movement with the fused QKV computation.
These effects provide the additional 7--9\(\times\) speedup over the per-row path in the main range.

\subsection{Fast Candidate Selection Kernel Evaluation}
\label{subsec:eval-e3-topk}

\begin{figure}[t]
  \centering
  \includegraphics[width=0.9\columnwidth]{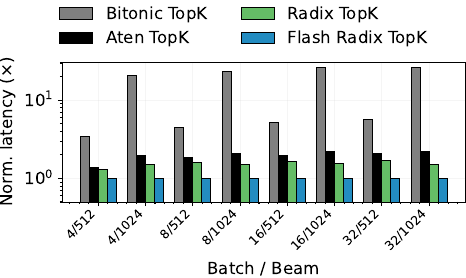}
  \vspace{-2mm}
  \caption{Beam-search Top-K latency for GR-style candidate selection with vocabulary size 8192.}
  \Description{A log-scale grouped bar chart comparing Top-K latency across request batch sizes 4, 8, 16, and 32 and beam widths 512 and 1024. \sys's shape-aware Top-K selection is the lowest bar in each group, while forced WarpBitonic becomes much slower at beam width 1024.}
  \label{fig:eval-e3-topk}
  \vspace{-2mm}
\end{figure}

We now isolate candidate selection to evaluate the beam-search-side performance contribution to \sys.
As shown in Figure~\ref{fig:eval-e3-topk}, under a SID vocabulary size of 8192,
\sys is the fastest implementation in every tested setting.
At batch 32 and beam width 512, shape-aware selection takes 0.618\,ms, giving 2.08\(\times\) over ATen and 1.70\(\times\) over standard radix.
At beam width 1024, the same batch size takes 1.307\,ms, giving 2.24\(\times\) over ATen and 1.50\(\times\) over standard radix.
Across the full sweep, \sys improves over ATen by 1.40--2.24\(\times\) and over standard radix by 1.32--1.70\(\times\).
The Bitonic path is especially mismatched once \(K\) reaches 1024, where it is 20.6--26.6\(\times\) slower than \sys.
The advantage comes from using a fast flash radix Top-K kernel for many 8192-wide per-beam rows, and from keeping the much wider request-level select on a separate multi-block radix path.
ATen and the standard radix baseline flatten this work into a generic Top-K shape, and the Bitonic path performs comparison work that scales poorly at large \(K\).

\subsection{Online Stability Validation}
\label{subsec:eval-e4-online-stability}

\begin{figure}[t]
  \centering
  \includegraphics[width=0.9\columnwidth]{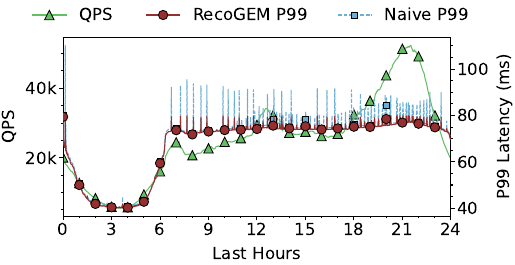}
  \vspace{-2mm}
  \caption{QPS and P99 latency over 24h.}
  \Description{A dual-axis line plot showing QPS and P99 latency over a 24h period.}
  \label{fig:24h}
  \vspace{-4mm}
\end{figure}

We next ask whether the runtime--engine decoupling architecture translates into stable runtime operability under live production traffic.
Over the full deployment history of \sys, the cluster has reached an all-time peak of 60{,}000 QPS while sustaining P99 latency below 80\,ms and an average MFU of 55.3\%. Figure~\ref{fig:24h} shows a representative 24-hour production trace. Traffic varies substantially during the day, from $\sim$5{,}000 QPS in low-traffic periods to more than 50{,}000 QPS at peak. Despite this variation, \sys keeps end-to-end P99 latency below 80\,ms throughout the trace.

Figure~\ref{fig:24h}  also highlights the effect of \sys's live-update path. The dashed \emph{Naive P99} curve corresponds to our earlier update implementation, which rebuilt the full execution engine and pushed the resulting artifact directly to serving instances during rollout. This approach made online updates interfere with active inference and produced repeated tail-latency spikes, with 2–3$\times$ bigger P99 latency spike during update periods. In contrast, \sys prepares and validates the next engine–contract pair outside the active request path, and cuts over by pointer switching to the validated version. As a result, frequent artifact updates are absorbed without visible SLA violations in the production trace. This result is important operationally: for GR serving, update handling is part of the latency-critical system design, not merely a deployment convenience.

%% file: sec/06-lessons.tex
\section{Discussion}
\label{sec:lessons}

\noindent\textbf{Evolution of the GR Serving System.}
\label{subsec:serving-evolution}
Given the architectural similarity between LLM and GR models, our first attempt was to adapt existing LLM inference frameworks for GR serving. This turned out to be the wrong systems boundary for two reasons. First, the required changes were invasive. Retrofitting an LLM engine to accommodate customized I/O formats, heterogeneous step transitions, online updates, as well as business logic across recommendation, search and advertising  required significantly more engineering effort than building a GR-native serving system from scratch.
Second, GR serving is a brownfield deployment problem. In our environment, the new serving system had to coexist with legacy RecSys infrastructure, reuse feature and embedding capabilities, and preserve compatibility with existing request paths.
These constraints made a monolithic adaptation of LLM stacks both costly and fragile. They instead favored the architecture used by \sys: a narrow execution engine surrounded by a GR-aware runtime that owns recommendation-specific responsibilities.

\noindent\textbf{Performance Bottlenecks Shift in GR.}
\label{subsec:performance-bottlenecks}
A key deployment lesson is that GR serving completely rewrites the performance overhead breakdown of RecSys. Unlike DLRM-style serving, where sparse embedding lookups and feature materialization dominate (often exceeding 50\% of total latency~\cite{wan2021flashembedding, naumov2019dlrm,gupta2020architectural}), embedding lookups in our GR deployment account for less than 10\% of the end-to-end time.
Instead, the computational weight shifts to the generative decoder.
Candidate selection is costly when executed with generic kernels, and systems like xGR~\cite{sun2025xgr} report that it takes over 40\% of serving time.
In our deployment, careful kernel-level and system optimizations reduce the candidate selection overhead to less than 10\% even for very large candidate sets, shifting performance pressure back to the dense decode execution.

\noindent\textbf{Execution Engine Generation.}
Yet another lesson was to move TensorRT engine generation out of the online deployment path. Our early prototypes used a two-stage conversion pipeline: the PyTorch model was exported offline to ONNX, an open model-interchange format, and the ONNX model was converted to a TensorRT engine during deployment. This design introduced two failure points. Export failures were at least isolated to the offline model build, but TensorRT conversion failures could surface during rollout, where they were harder to debug across boundaries. \sys instead builds TensorRT engines directly in the offline release pipeline and publishes only validated engine–contract pairs to serving. In our environment, moving engine generation offline reduced the release-side deployment time by 40\%. This is critical for online services, as minimizing downtime during an operational error prevents massive revenue losses that can accumulate by the second. More importantly, it made step profiles, plugin paths, and quantization choices explicit build-time decisions, so each choice can be tested with the engine artifact before the version reaches serving cutover.

\noindent\textbf{Generality.}
GR is a widely deployed workload, running in production for recommendation, search, advertising, and e-commerce~\cite{deng2025onerec,zhou2025onerecv2,han2025mtgr,agarwal2025pinrec,xue2026gr4ad,chen2025unisearch,zhang2026onemall}.
\sys builds on its workload structure rather than on recommendation semantics: a few decoding steps over many short candidate sequences, with per-step selection over a predefined vocabulary.
Its optimizations like attention kernels and candidate selection therefore apply under both large batches and wide beams, and carry over to generative retrieval~\cite{tay2022dsi}.
They may further extend to the recently released Jev~\cite{typesafe2026jev}, where each request evaluates many independent questions against one shared input, each over a predefined option set.

\noindent\textbf{Scalability and Future Directions.}
\label{subsec:future-directions}
Since requests share no state across GPUs, \sys's throughput scales near-linearly with the GPU count.
As GR model scale and architectural complexity continue to grow, however, a natural next step is distributed GR serving.
Tensor and pipeline parallelism should be integrated with step-granular execution, rather than layered on top as opaque model-parallel wrappers. For sparse GR models, expert parallelism will likely become equally important: the serving contract must eventually describe cross-rank state, expert placement, routing constraints, and version-consistent cutover across distributed artifacts.
Beyond scaling out, \sys also opens opportunities for step-aware model compression. Existing quantization techniques show that precision can be treated as an important deployment knob; GR adds a new dimension because different SID steps may carry different semantic importance and expose different execution shapes. Future systems can exploit this structure to choose per-step precision or quantization profiles, improving hardware efficiency while preserving recommendation quality.

%% file: sec/07-related-work.tex
\section{Related Work}
\label{sec:related-work}

\noindent\textbf{GR Serving Systems.}
\label{subsec:related-gr-serving}
Existing GR serving systems optimize within an inherited worldview rather than reconsidering the serving abstraction itself~\cite{sun2025xgr,guo2025flame}.
xGR and FLAME each commit to a fixed substrate: xGR treats GR as a long-prompt, short-output variant of LLM serving and accelerates individual components~\cite{sun2025xgr}, while FLAME casts GR as a TensorRT deployment problem and disaggregates CPU feature preprocessing from GPU model execution~\cite{guo2025flame}.
Both tune components within the inherited substrate and retain a homogeneous decode loop, leaving the step heterogeneity unexploited.
A separate line of work advances GR on the algorithm side: MTGR scales GR model architectures to industrial deployment~\cite{han2025mtgr}, GR4AD brings GR to large-scale advertising with lazy autoregressive decoding and dynamic beam strategies~\cite{xue2026gr4ad}, and speculative decoding accelerates beam-search generation itself~\cite{lin2025efficient}. These efforts redesign the model and its decoding procedure rather than the serving system.
\sys differs from existing efforts
as it derives its serving abstraction from GR's workload objectives, exposes SID step boundaries, and co-designs a GR-serving runtime with a stepwise optimized execution engine.

\noindent\textbf{LLM and RecSys Serving Systems.}
\label{subsec:related-adjacent-serving}
GR workloads fall between two mature serving stacks.
LLM serving organizes execution around a homogeneous next-token loop over KV-centric state, from iteration-level scheduling and paged KV caching to engine-level decode optimization~\cite{yu2022orca,kwon2023pagedattention,agrawal2024taming,zheng2024sglang,nvidia2026tensorrtllm}, with attention and selection kernels tuned for long, homogeneous decoding iterations~\cite{dao2022flashattention,ye2025flashinfer,li2024radik, Gao2025WeaverEM}; these assumptions do not hold for GR's candidate state and its short, beam-expanded, heterogeneous decode steps.
RecSys serving contributes the operational discipline GR inherits---heterogeneous feature materialization and frequent online model refresh~\cite{naumov2019dlrm,gupta2020architectural,xie2020kraken,sima2022ekko,yang2025prism}---but its pipelines end at a scoring boundary and carry no generation state across steps.
\sys bridges the two, keeping the LLM execution substrate for computation while promoting the recommendation stack's candidate selection and online freshness requirements into serving contract obligations.

%% file: sec/08-conclusion.tex
\section{Conclusion}
\label{sec:conclusion}
The recent paradigm shift towards SID-based GR introduces complex recommendation serving demands that traditional LLM and RecSys infrastructures cannot efficiently handle.
To accommodate GR-specific architectural constraints and optimization opportunities, we propose \sys, an in-production GR serving system that uses an explicit deployment contract to manage heterogeneous step transitions and online updates outside the compute engine. Additionally, \sys introduces step-granular execution planning combined with specialized decode operators, namely row-tiled self-attention and fast candidate selection, to exploit GR-specific execution shapes. Evaluations on our production-scale GR workloads demonstrate that \sys successfully mitigates performance bottlenecks while sustaining runtime operability, achieving high throughput and strict tail-latency compliance across various real-world scenarios.